\documentclass[twoside]{article}
\usepackage{Proc_NTSE_14Prime}
\usepackage{color}
\begin{document}
\thispagestyle{plain}
\publref{Vary}

\begin{center}
{\Large \bf \strut
{\itshape Ab Initio} No Core Shell Model --- Recent Results\\ and Further Prospects\label{JPV}
 \strut}\\
\vspace{10mm}
{\large \bf 
James P. Vary$^{a}$, Pieter Maris$^{a}$, Hugh Potter$^{a}$, \strut\\
Mark A. Caprio$^b$, 
Robin Smith{$^{c}$},
Sven Binder{$^{d,e}$},
\strut Angelo Calci{$^{f}$},
Sebastian Fischer{$^{g}$},
\strut Joachim~Langhammer{$^{g}$},
Robert Roth{$^{g}$},
Hasan~Metin~Aktulga{$^{h,i}$}
\strut Esmond~Ng$^i$, Chao~Yang$^i$, Dossay~Oryspayev$^{j}$, Masha~Sosonkina$^k$,
Erik Saule$^{l}$\\
and \"{U}mit~\c{C}ataly\"{u}rek$^{m,n}$.\strut
}
\end{center}

\noindent{
\small $^a$\it Department of Physics and Astronomy, Iowa State University, Ames, IA   50011,  USA} \\
{\small $^b$\it Department of Physics, University of Notre Dame, Notre Dame, IN 46556, USA} \\
{\small $^c$\it School of Physics and Astronomy, University of Birmingham, Birmingham B15 2TT, UK}\\
{\small $^d$\it Department of Physics and Astronomy, University of Tennessee, Knoxville, TN 37996, \phantom{$^d$}USA}\\
{\small $^e$\it Physics Division, Oak Ridge National Laboratory, Oak Ridge, TN 37831, USA} \\
{\small $^f$\it TRIUMF, 4004 Wesbrook Mall, Vancouver, British Columbia, V6T 2A3, Canada}\\
{\small $^g$\it Institut f\"ur Kernphysik, Technische Universit\"at Darmstadt, 64289 Darmstadt, Germany}\\
{\small $^h$\it Department of Computer Science, Michigan State University, East Lansing, MI 48844, \phantom{$^h$}USA} \\
{\small $^i$\it Lawrence Berkeley National Laboratory, Berkeley, CA 94720, USA}\\
{\small $^j$\it Department of Electrical and Computer Engineering, Iowa State University, Ames, IA   \phantom{$^d$}50011,  USA}\\
{\small $^k$\it Department of Modeling, Simulation and Visualization Engineering, Old Dominion Univer- \phantom{$^d$}sity,
  Norfolk, VA   23529, USA}\\
{\small $^l$\it Department of Computer Science, University of North Carolina Charlotte, Charlotte, NC \phantom{$^l$}28223, USA}\\
{\small $^m$\it Department of Biomedical Informatics, The Ohio State University, Columbus, OH 43210, \phantom{$^m$}USA}\\
{\small $^n$\it Department of Electrical and Computer Engineering, The Ohio State University, Columbus, \phantom{$^d$}OH 43210, USA}

\markboth{
James P. Vary {\it et al.}}
{
{\it Ab Initio} No Core Shell Model} 

\begin{abstract}
There has been significant recent progress in solving the long-standing problems of how nuclear shell structure and collective motion emerge from underlying microscopic inter-nucleon interactions.  We review a selection of recent significant results within the {\it ab initio} No Core Shell Model (NCSM) closely tied to three major factors enabling this progress: (1) improved nuclear interactions that accurately describe the experimental two-nucleon and three-nucleon interaction data; (2) advances in algorithms to simulate the quantum many-body problem with strong interactions; and 
 (3) continued rapid development of high-performance computers now capable of performing $20 \times 10^{15}$ floating point operations per second. 
We also comment on prospects for further developments.
\\[\baselineskip] 
\end{abstract}

\section{Introduction}




The {\it ab initio} No Core Shell Model (NCSM), using realistic microscopic nucleon-nucleon ($NN$) and three-nucleon forces ($3N$Fs), has proven to be a powerful combination for describing and predicting properties of light nuclei \cite{Navratil:2000ww,Navratil:2000gs,Navratil:2007we,Maris09_NCFC,Barrett:2013nh,Maris:2013poa,Shirokov_review_JPV382:2014}. The Hamiltonian framework results in a large sparse matrix 
eigenvalue problem for which we seek the low-lying eigenvalues and eigenvectors to form comparisons with experimental data and to make testable predictions. Given the rapid advances in hardware with frequent 
disruptions in architecture, it has become essential for physicists, computer scientists and
applied mathematicians to work in close collaboration in order to achieve efficient solutions 
to forefront physics problems.
Fortunately, US funding agencies have recognized these challenges at the interface of science and technology and have provided support leading to our recent successes 
\cite{Sternberg:2008,Vary:2009qp,Maris_JPV:2010,Aktulga_JPV324:2011,Maris:2012du,Aktulga_JPV337:2012,Aktulga_JPV366:2013,Potter:2014gwa,Potter:2014dwa,Yang_JPV379:2014}. 

We present here a selection of recent results for light nuclei and neutron drops in external traps 
and set out some of the challenges that lie ahead.  The results include both those utilizing the JISP16 $NN$ interaction and those using chiral effective field theory $NN$ plus $3N$ interactions. We also present a selection of algorithms developed for high-performance computers that are helping to rapidly pave the way to efficient utilization of exascale machines ($10^{18}$ floating point operations per second).  We illustrate the scientific progress attained with multi-disciplinary teams of physicists, computer scientists and applied mathematicians. 

This paper is aimed to complement presentations at this meeting that cover closely-related topics.  In this connection, it is important to point especially to the papers by Dytrych {\em  et al.}~\cite{Dytrych_here}, 
by Abe {\em et al.}~\cite{Abe_here}, 
by Shirokov {\em et al.}~\cite{Shirokov_here} 
and by  Mazur {\em et al.}~\cite{Mazur_here}.  We therefore focus here on the following recent results: 
(1) demonstrating the emergence of collective rotations in light nuclei; 
(2) achieving an accurate description of the properties of $^{12}$C with chiral Hamiltonians; 
(3) solving for properties of neutron drops with chiral Hamiltonians; 
(4) development of techniques for efficient use of computational accelerators; and 
(5) development of techniques for overlapping communication and computation.


\section{Emergence of collective rotations}

NCSM calculations of various types have been used to demonstrate the emergence of collective rotational correlations in $p$-shell nuclei, including $^6$Li \cite{Dytrych:2013cca,Dytrych_here}, the Be isotopes \cite{Dytrych:2013cca,Caprio:2013yp,Maris:2014jha,Caprio:2014_JPV}, and $^{12}$C 
\cite{Maris:2013a}.
Here we focus on the results for the Be isotopes solved in the 
    No Core Full Configuration (NCFC) 
     framework 
\cite{Maris09_NCFC, Maris:2013poa,Shirokov_review_JPV382:2014} using
the realistic JISP16 $NN$ interaction \cite{Shirokov:2004ff,Shirokov:2005bk} with 
the $M$-scheme harmonic oscillator (HO) basis. The NCFC framework 
uses many of the same techniques as the NCSM but additionally 
features extrapolations of observables to the 
infinite matrix limit \cite{Maris09_NCFC}.

With no prior selection of our 
 basis to favor solutions with collective motion 
and using only the realistic bare $NN$ interaction (i.\;e. we omit the Coulomb interaction to ensure exact conservation of isospin thereby simplifying the spectrum\footnote{%
The primary effect
of the Coulomb is to shift the binding energies which would not affect our analysis of rotational band observables.  New analysis including Coulomb~\cite{MPVSIJMPT} 
confirms this.})
we face the task 
of analyzing our microscopic results and determining which particular states, 
among the large number of calculated levels, exhibit signatures of collective nuclear motion.  
We follow the path of calculating observables 
and post-analyzing their systematics to infer that they follow the patterns prescribed
by collective rotation.  This path is analogous to that taken when analyzing
experimental data. When we discover patterns appropriate to a collective band 
in our calculated results, we assign the moniker of ``collective motion'' to our
microscopic results.  We further compare the so-detected band with experimental
results and find good agreement which further supports our discovery of emergent 
collective phenomena in light nuclei from the underlying microscopic many-body theory.

The details of this step-by-step analysis may be found in the Refs. 
\cite{Caprio:2013yp,Maris:2014jha,Caprio:2014_JPV}.  We analyze 
the systematics of calculated excitation energies, quadrupole moments, dipole moments, 
electric quadrupole transition $B(E2)$'s and their reduced matrix elements
to isolate states which have a clear rotation band assignment from those which do not.
In this way, we have identified both ground state and excited state bands, both natural and unnatural 
parity bands, and bands in even-even as well as in even-odd nuclei.  

Perhaps the most striking hallmark of collective rotation is the appearance of excited states with excitation energies that follow a simple pattern prescribed by the collective model. 
This pattern of collective rotational 
excitation energies is given in Eq.~\eqref{Collective_excitation_energy}: 
\begin{equation}
E(J)=E_0+A\bigl[J(J+1)+ a(-)^{J+1/2}(J+\tfrac12)\delta_{K,1/2} \bigr],
\label{Collective_excitation_energy}
\end{equation}
where $E_0$ is an offset to properly position excited band heads 
relative to the lowest band head, $a$ is the Coriolis decoupling parameter for $K=\frac{1}{2}$ bands appearing in \mbox{odd-$A$} nuclei,
$J$ is the total angular momentum
and $A\equiv\hbar^2/(2\cal{J})$  with $\cal{J}$ representing 
the moment of inertia of the deformed nucleus.

To be convinced that the states are indeed members of a rotational band 
one needs to find that these states also exhibit enhanced electromagnetic 
moments and transition rates that exhibit a dependence on 
angular momentum $J$ that is also prescribed by the collective rotational
model. We therefore adopt these additional criteria for assigning calculated
states to rotational bands. It is worth noting here that, in light nuclei, gamma decay data are scarce 
due to the short-lived resonant nature of the states. Therefore, the calculations provide 
access to quantities that are typically inaccessible in experiment, yet crucial for confirming collectivity.

\begin{figure}[!t]
\centerline{\includegraphics[width=1.0\textwidth]{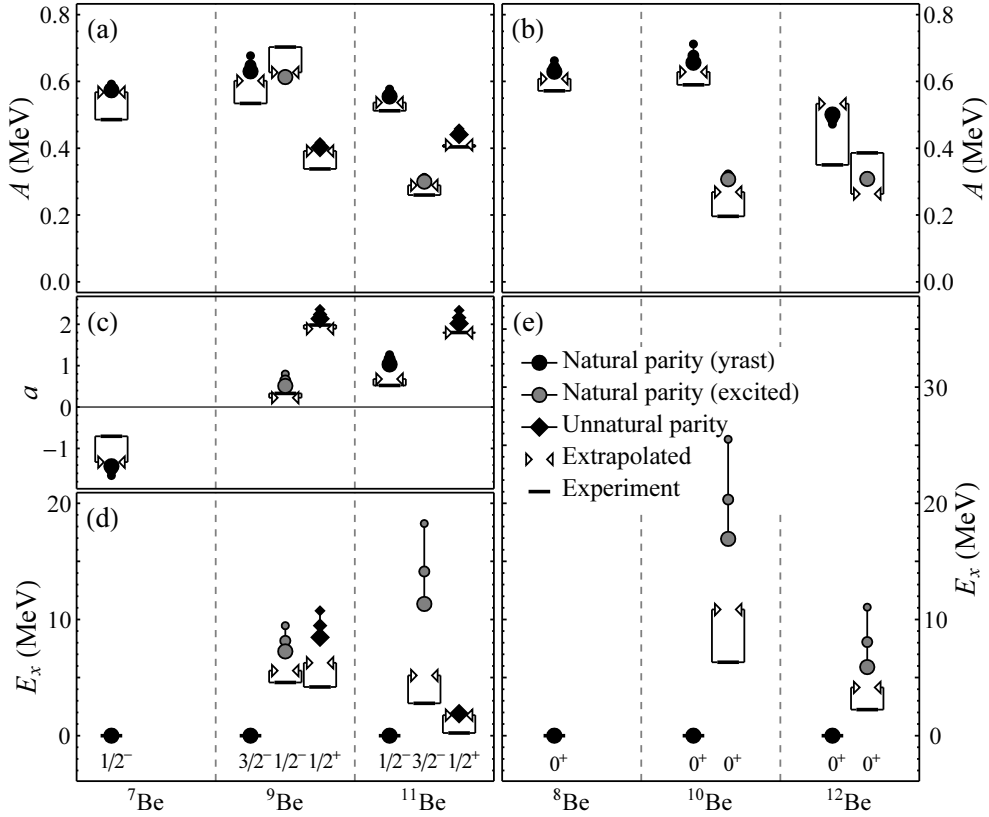}}
\caption{Rotational parameters $A$, $a$ and $E_x$ [defined relative to the yrast band
as~$E_x\equiv E_0-E_{0,\text{yrast}}$~--- see Eq.~\eqref{Collective_excitation_energy}]
for ground and excited bands of the Be isotopes (adapted
from Ref~\cite{Caprio:2014_JPV}). Brackets highlight the difference between 
the parameters determined from
experimental data (horizontal bars) and those extracted from NCFC calculations with
extrapolation (parallel triangles) to the infinite matrix limit.  Solid symbols connected 
by solid lines indicate the finite matrix results as a function of increasing $N_{\rm max}$ 
with larger symbols for larger $N_{\rm max}$ values. $N_{\rm max}$ is defined as the maximum number of oscillator quanta in the HO configurations above the minimum for the
nucleus under investigation. The minimum $N_{\rm max}$ is $0$ for natural
parity and~$1$ for unnatural parity. The results indicated in the solid 
symbols correspond to~${6\leq N_{\rm max} \leq10}$
for natural parity and $7\leq N_{\rm max} \leq11$ for unnatural parity.}
\label{Collective_parameters_Be}      
\end{figure}

We extract parameters 
of the traditional rotational description through fits to our theoretical results 
after extrapolation to the the infinite matrix limit (for extrapolation details
see Ref.~\cite{Caprio:2014_JPV})
and we compare these extracted parameters with rotational parameters determined from 
similar fits to the corresponding experimental data.
The energy parameters for bands across the
Be isotopic chain are summarized in Fig.~\ref{Collective_parameters_Be}:
the band excitation energy $E_x$ (defined relative to the yrast band
as $E_x\equiv E_0-E_{0,\text{yrast}}$), the band
rotational parameter or slope $A$, and the band Coriolis decoupling
parameter or staggering $a$ (for $K=1/2$). 

In total, we compare 23 theoretical and experimental collective rotation parameters for
energies in the 6 Be isotopes depicted in Fig.~\ref{Collective_parameters_Be}. 
Overall the agreement
between theory and experiment is remarkable.  Additional analyses of the 
calculated electromagnetic observables 
in Refs. \cite{Caprio:2013yp,Maris:2014jha,Caprio:2014_JPV} and comparison with
sparse data available confirm that
we have observed the emergent phenomena of collective rotation in these {\it ab initio} 
calculations for the Be isotopes.  At the same time, there are opportunities for additional theoretical 
and experimental research to explore, for example, where rotational bands terminate 
and whether additional bands may be found in these and other light nuclei.  It
appears that bands do not always terminate at the state corresponding to the
maximum angular momentum supported by the nucleons occupying the standard
valence shell model orbitals \cite{Caprio:2013yp,Maris:2014jha,Caprio:2014_JPV}.

\section{Chiral Hamiltonian description of \boldmath$^{12}$C}

Recent significant theoretical advances for the underlying Hamiltonians, constructed within chiral effective field theory (EFT), provide a foundation for nuclear many-body calculations rooted in QCD \cite{EpHa09,MaEn11}. These developments motivate us to adopt a chiral EFT Hamiltonian 
here and in the following section on neutron drops in an external trap.
We also adopt the similarity renormalization group (SRG) approach \cite{Glazek:1993rc,Wegner:1994,Bogner:2007rx,Hergert:2007wp,Bogner:2009bt,Furnstahl:2012fn} that allows us to consistently evolve (soften) the Hamiltonian and other operators, including $3N$ 
interactions~\cite{Roth:2011ar,Jurgenson:2009qs,Roth:2013fqa}.

We select the example of the spectroscopy of $^{12}$C to illustrate the recent progress.
In so doing, it is important to note that additional progress in achieving larger basis
spaces is needed before we can realistically address cluster model states in light
nuclei such as the celebrated ``Hoyle state'', a $0^+$ state at 7.654 MeV of excitation energy
in~$^{12}$C.

\begin{figure}[!t]
{\centerline{\includegraphics[width=0.74\textwidth]{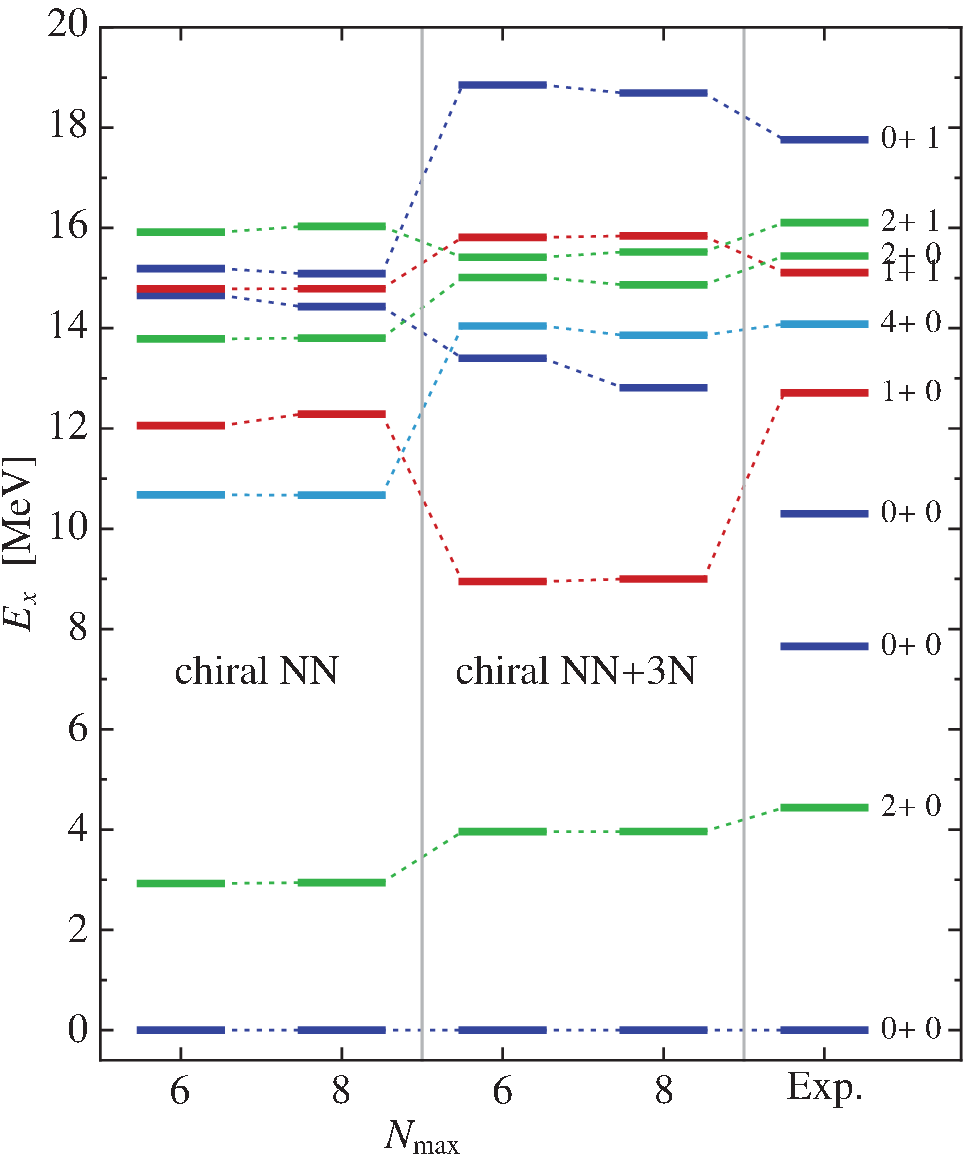}}\vspace{2.8ex}
\centerline{\includegraphics[width=0.74\textwidth]{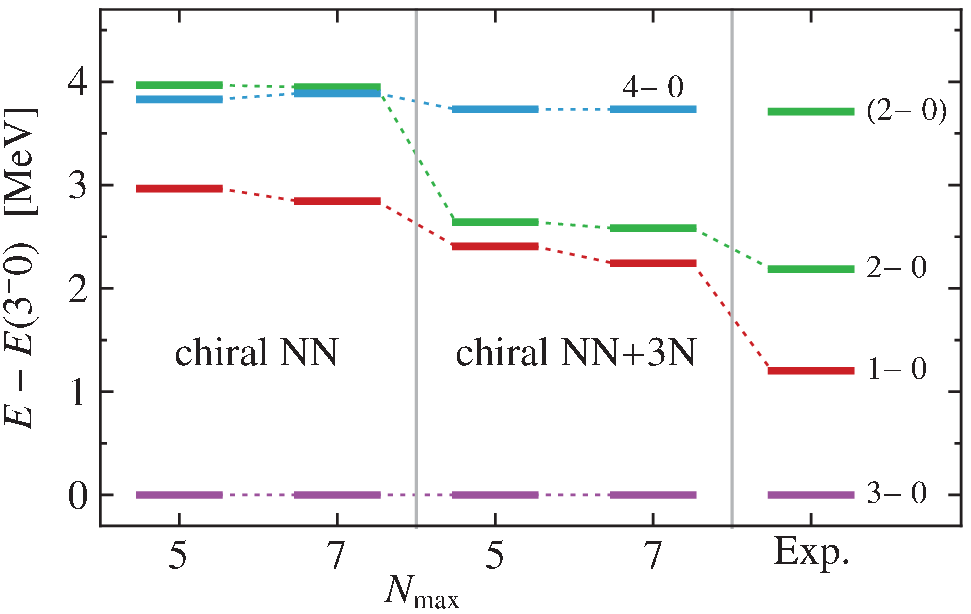}}}\vspace{-.3ex}
{\caption{Theoretical and experimental excitation spectra of $^{12}$C for both
positive parity (top panel) and negative parity (bottom panel) states for two different
values of $N_{\rm max}$ at $\hbar\Omega=20$ MeV (adapted from 
Ref.  \cite{Maris:2014_JPVxxx}). The columns 
labelled ``chiral $NN$'' include the $3N$F induced by SRG while the sub panels labelled
``chiral $NN + 3N$'' include the initial $NN + 3N$F evolved by SRG together with $NN$.  The SRG
evolution parameter is $\lambda = 2.0$ fm$^{-1}$. See Ref. \cite{Maris:2014_JPVxxx} 
for additional details.}\vspace{-2ex}
\label{C12_excitation_spectra}}   
\end{figure}

The theoretical excitation spectra are presented in Fig.~\ref{C12_excitation_spectra} for the two highest 
$N_{\rm max}$ values currently achievable and are compared with experiment. 
For the negative parity states, we elect to show excitation energies relative to the 
lowest state of that parity whose experimental energy is 9.641 MeV above the ground state.
The trends with increasing $N_{\rm max}$
(see the trends for additional observables in Ref. \cite{Maris:2014_JPVxxx}) 
suggest convergence is sufficient to draw important conclusions regarding
the underlying interaction. In particular, we note that the shifts from including the
initial $3N$ interaction are substantial.  In most cases, these shifts improve agreement
between theory and experiment.  A notable exception is the $J^{\pi}=1^+$, $T=0$ positive
parity state which shifts further from experiment when we include 
the initial $3N$ interaction.  

From our results in $^{12}$C, we conclude that we need further improvements in the chiral interactions.
For example, we need to have $NN$ and $3N$ interactions at the same chiral order to be consistent.  We also
need to extend the chiral order of the interactions to N4LO and, possibly, to include the derived 
four-nucleon ($4N$) interactions.

\section{Confined neutron drops with chiral Hamiltonians}

There are many motivations for considering artificial pure neutron systems 
confined by an external trap.  

\begin{itemize}
\item{Gain insights into the properties of systems dominated by multi-neutron degrees of freedom such as unstable neutron-rich nuclei and neutron stars.}
\item{Isolate selected isospin components of the $NN$ ($T=1$) and $3N$ ($T=3/2$) interactions for detailed study.}
\item{Inform the development of nuclear energy density functionals that may be tuned to reproduce {\it ab initio} calculations, complementing their tuning to experimental data.}
\end{itemize}

The external trap is required since realistic interactions do not bind pure neutron systems, 
though they do produce net attraction when the systems are confined.  
The main foci are to observe differences among realistic interactions and to see if subshell closures are predicted.  For example, one may investigate spin-orbit splitting as a function of the chosen interaction and as a function of the external field parameters.
%

%
\begin{figure}[!b]
\centerline{\includegraphics[width=.98\textwidth]{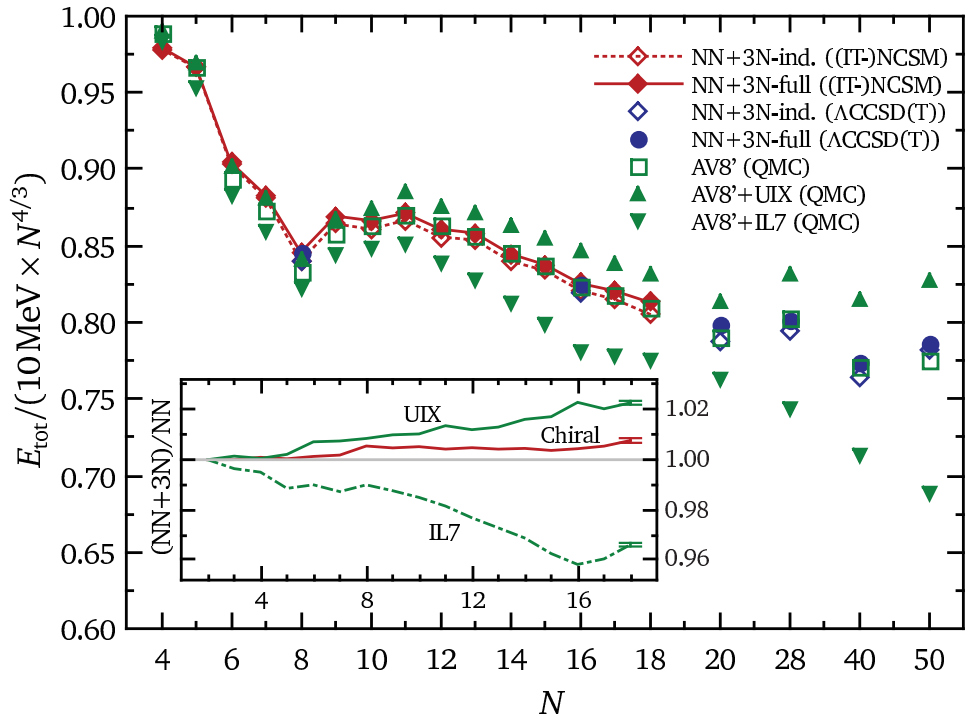}}
\caption{Comparison of ground state energies of systems with $N$ neutrons trapped 
in a 
    HO with strength 10 MeV.  Solid red diamonds and blue dots 
signify results with $NN+3N$ interactions derived from chiral effective field theory related to QCD. 
The inset displays the ratio of $NN+3N$ to $NN$ alone for the different interactions with the 
error indicated on the far right of each curve where it is maximum.
The label indicates the many-body methods employed: 
(Importance-Truncated) No Core Shell Model ((IT-)NCSM);
Coupled Cluster including Triples ($\Lambda$CCSD(T));
Quantum Monte Carlo (QMC). Figure adapted from Ref. \cite{Potter:2014dwa}.
}
\label{Neutron_drops}      
\end{figure}

Using the same realistic chiral $NN + 3N$ interactions as used in the previous section, we investigated
\cite{Potter:2014gwa,Potter:2014dwa} neutron drop systems in a 10 MeV HO 
trap.
In Ref. \cite{Potter:2014dwa} we compared the results with those from 
Green's Function Monte Carlo (GFMC) and auxiliary
field diffusion Monte Carlo (AFDMC) \cite{Gandolfi:2010za,Maris:2013rgq}
using the Argonne $v_8'$ (AV8') $NN$ interaction 
\cite{Pudliner:1997ck} and the Urbana IX (UIX) $3N$ interaction. 
We also compared with GFMC and AFDMC results using
AV8' with the Illinois-7 (IL7) $3N$ interaction \cite{Pudliner:1997ck,Pieper:2008}.

For the investigations in Ref. \cite{Potter:2014dwa} we employed both NCFC and coupled cluster (CC) 
methods.  By implementing CC, we were able to obtain results for larger neutron drop systems.

We found important dependences on the selected interactions as shown 
in Fig.~\ref{Neutron_drops} which should have an impact 
on phenomenological energy-density functionals that may be derived from them.  
Note in Fig.~\ref{Neutron_drops} that, with increasing $N$, the chiral predictions 
lie between results from different high-precision phenomenological interactions, 
i.\;e. between AV8$^\prime$+UIX and AV8$^\prime$+IL7.  It will be very important to see the 
influences the results of these different interactions have on energy density functionals.

One also notices in Fig.~\ref{Neutron_drops} there are surprisingly weak contributions 
from the inclusion of the chiral $3N$ interaction. Based on systematic trends shown 
in previous neutron-drop investigations \cite{Gandolfi:2010za,Maris:2013rgq,Bogner:2011kp}, 
with non-chiral interactions we anticipate these conclusions will persist 
over a range of HO well strengths.  Additional investigations are in progress 
to confirm this hypothesis and to extend the results to higher neutron numbers.

\section{Computational accelerators\\ and decoupling transformations}

Fundamental physics investigations with chiral $NN+3N$ interactions require forefront computational
techniques in order to efficiently utilize leadership computational facilities.  Many of our efforts 
are aimed to develop new algorithms that exploit the recent advances in hardware and software.
Here we describe one of those projects that could only have been accomplished through our
multidisciplinary team working in close collaboration.

This specific project focused on adapting our NCSM code, Many-Fermion Dynamics --- nuclear (MFDn), for use with GPU accelerators on the supercomputer Titan at Oak Ridge National Lab.  MFDn represents the input $NN$ and $3N$ interactions in the ``coupled-$JT$'' basis with coupled angular momentum and isospin, exploiting rotational symmetry and isospin conservation to reduce memory requirements \cite{Roth:2011ar,Roth:2013fqa,Maris:2013a}.  In one representative case, storing a $3N$ input interaction in the coupled-$JT$ basis reduces the interaction file size from 33 GigaBytes (GB) to less than 0.5 GB.  This method is crucial for pushing the boundaries of problem sizes that we can address, as the input interactions must be stored once per process; using the ideal process configuration on Titan, processes have access to 16 GB each.  Such a reduction in memory usage, then, not only enables calculations with larger input interactions, which are required for larger model spaces, but also makes their memory footprints more manageable, leaving more room for the memory-limited NCSM calculation.

\begin{figure}[t!]
\centerline{\includegraphics[width=.72\textwidth]{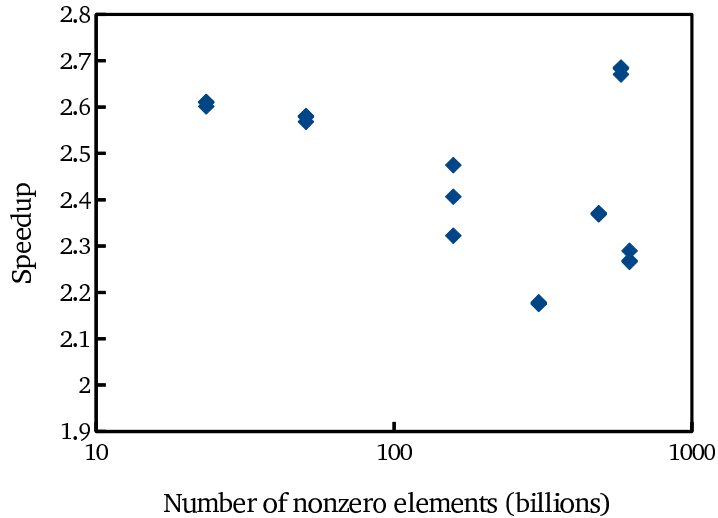}}
\caption{Speedup in the many-nucleon Hamiltonian construction stage due to implementation on GPU accelerators, graphed against the number of nonzero matrix elements in the Hamiltonian.  There is no clear trend, but all speedups are in approximately the same region, indicating good weak scaling across this range of problem sizes.  We graph matrix construction speedup here instead of overall speedup; overall speedup depends strongly on how long the matrix diagonalization takes, which is a function of the number of eigenstates required.  Figure adapted from Ref. \cite{Potter:2014gwa}.
}
\label{GPU_speedup}      
\end{figure}

As a side-effect of this compression, as we construct the full many-nucleon Hamiltonian from the input $NN$ and $3N$ interactions, we must perform basis transformations to extract input interaction matrix elements that our code can use directly.  These basis transformations are both computationally intensive and amenable to parallelization; they are a natural fit for Titan's GPU accelerators.  We have taken advantage of our multidisciplinary team of physicists, computer scientists, and applied mathematicians to port this section of our code to the GPU and optimize it\cite{Oryspayev:2013}.  Integrating the GPU-accelerated basis transformation into MFDn produces a speedup of 2.2x--2.7x in the many-nucleon Hamiltonian construction, as illustrated in Fig.~\ref{GPU_speedup}, and a speedup of 1.2x--1.4x in the full calculation, with some variation depending on the particular problem chosen \cite{Potter:2014gwa}.

\section{Overlapping communications and calculations}

Our configuration interaction (CI) approach to the nuclear many-body problem results in a large 
sparse matrix eigenvalue problem with a symmetric real Hamiltonian matrix.  This presents
major technical challenges and is widely recognized as ``computationally hard.''  One of the 
popular methods for obtaining the low-lying eigenvalues and eigenvectors is the Lanczos algorithm 
that we have implemented in MFDn.   As the problem size increases with either increasing basis
spaces or with the inclusion of $3N$ interactions, we face the challenge of communication costs rising with
the increased numbers of nodes used in the calculations.  The increase in nodes is driven by memory
requirements as mentioned in the previous section.  

In order to reduce communication costs, we developed an efficient mapping of the eigensolver onto the available hardware with a ``topology-aware'' mapping 
algorithm~\cite{Aktulga_JPV337:2012,Yang_JPV379:2014}. We also developed an improved Lanczos 
algorithm that overlaps communications with calculations~\cite{Aktulga_JPV366:2013,Yang_JPV379:2014}.

For the challenge of efficiently overlapping communications with calculations, 
we worked with a hybrid MPI-OpenMP
implementation and delegated one or a few threads to perform inter-process communication tasks, while the remaining threads carried out the multi-threaded computational tasks.  In our algorithm, we also implemented a dynamical scheduling of the computations among the threads for the sparse 
matrix-vector multiplication (SpMV) so that, once a communication thread completes that task, 
it can participate in the multi-threaded computations.

\begin{figure}
\includegraphics[width=1.00\textwidth]{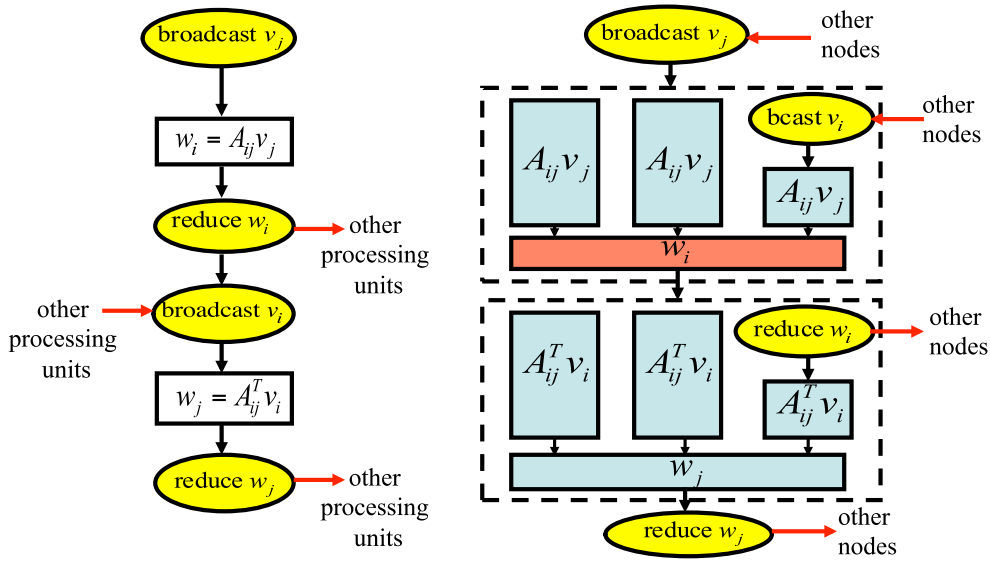}
\caption{Comparison of 
      SpMV 
       and communication methods
for an iteration of the Lanczos algorithm carried out by the majority of the processing units, the ones
that store the off-diagonal blocks of the Hamiltonian matrix.
The left subfigure displays a traditional sequential process that may be implemented with MPI.  The 
right subfigure presents our algorithm suitable for hybrid MPI-OpenMP.  Yellow ovals depict communication
and rectangles depict computation.  The red rectangle indicates where we require thread synchronization 
which incurs a small additional cost.
The figure is adopted from Refs.~\cite{Aktulga_JPV366:2013,Yang_JPV379:2014}.
}
\label{SpMV_overlap}      
\end{figure}

In Fig.~\ref{SpMV_overlap} we compare a straightforward SpMV implementation using sequential steps (left subfigure) with our algorithm (right subfigure).  By mapping MPI processes in a balanced column-major order as well as developing and implementing our algorithm to overlap communications and calculations, we achieved over 80\% parallel efficiency through reduction in communication overhead during the Lanczos iteration process.  This includes both the SpMV and orthogonalization steps that occur in each iteration.  We also found major improvements in the scalability of the \mbox{eigensolver} especially after adopting our topology-aware mapping algorithm.  Since SpMV and vector-vector multiplication of these types are common to many other iterative methods, we believe our achievements have a wide range of applicability.

\section{Future prospects}

Most of our applications have focused on light nuclei with atomic number $A \leq 16$ where our
theoretical many-body methods have achieved successes with leadership class facilities.
However, the frontiers of our field include applications to heavier nuclei and 
utilizing new and improved
interactions from chiral effective field theory.  At the same time, we aim to evaluate 
observables with increasing sophistication using their operators also derived within
chiral effective field theory.  We mention the example of neutrinoless double beta decay
as one exciting example of frontier research with {\it ab initio} computational nuclear theory.

We therefore face the dual challenge of advancing the underlying
theory at the same time as advancing the algorithms to keep pace with the growth
in the size and complexity of leadership class computers.  Recent history in these
efforts, with the substantial support of the funding agencies,
indicates we are experiencing a ``Double Moore's Law" rate of improvement~--- i.\;e.
Moore's Law for hardware improvements and a simultaneous Moore's Law improvement in
the algorithms/software.  
We need continued support for multi-disciplinary collaborations 
and growth in leadership class facilities in order to achieve 
the full discovery potential of computational nuclear physics.

\section{Acknowledgements}

This work was supported in part by the US National Science Foundation
under Grant No.\ PHY--0904782, the US Department of
Energy (DOE) under Grant Nos.~DE-FG02-87ER40371, DESC0008485
(SciDAC-3/NUCLEI) and DE-FG02-95ER-40934, by the Deutsche Forschungsgemeinschaft through contract SFB 634, by the Helmholtz International Center for FAIR (HIC for FAIR) within the LOEWE program of the State of Hesse, and the BMBF through contract 06DA7047I. 
This work was supported partially through GAUSTEQ 
(Germany and U.S. Nuclear Theory Exchange Program for QCD Studies 
of Hadrons and Nuclei) under contract number DE-SC0006758. 
A portion of the computational resources were
provided by the National Energy Research Scientific Computing Center
(NERSC), which is supported by the US DOE Office of Science, and by 
an INCITE award, ``Nuclear Structure and Nuclear Reactions'', from the US DOE
Office of Advanced Scientific Computing.  This research also used
resources of the Oak Ridge Leadership Computing Facility at ORNL,
which is supported by the US DOE Office of Science under Contract
DE-AC05-00OR22725. This research also used resources of the Argonne Leadership Computing Facility, 
which is a DOE Office of Science User Facility supported under Contract DE-AC02-06CH11357.
Further resources were provided by the computing 
center of the TU Darmstadt (Lichtenberg), the J\"ulich Supercomputing Centre (Juropa), and the LOEWE-CSC Frankfurt.

\end{document}